\documentclass[11pt]{interact}

\usepackage{amsmath}
\usepackage{amsfonts}
\usepackage{amsthm}
\usepackage{bm}
\usepackage{dsfont}
\usepackage{mathtools}
\usepackage{subcaption}
\usepackage{booktabs}
\usepackage{longtable}
\usepackage[usenames,dvipsnames]{color}

\usepackage{epstopdf}

\usepackage[numbers,sort&compress]{natbib}
\bibpunct[, ]{[}{]}{,}{n}{,}{,}
\usepackage{hyperref}
\hypersetup{
    colorlinks = true,
    citecolor = {blue},
}
\usepackage{hypernat}

\theoremstyle{plain}

\theoremstyle{definition}

\theoremstyle{remark}

\DeclareMathOperator*{\argmin}{arg\,min}

\newcommand{\hs}{\Omega}
\newcommand{\nhs}{\Omega^c}
\newcommand{\ns}{\Psi}
\newcommand{\vt}[1]{\boldsymbol{\theta}_{#1}}
\newcommand{\hvt}[1]{\hat{\boldsymbol{\theta}}_{#1}}
\newcommand{\cod}{\mathcal{T}}
\newcommand{\curv}{\mathcal{Z}}
\newcommand{\prob}{\mathbb{P}}
\newcommand{\expect}{\mathbb{E}}

\newcommand{\af}[1]{}
\newcommand{\rl}[1]{}
\newcommand{\jdt}[1]{}
\newcommand{\jwb}[1]{}

\begin{document}


\title{A Deterministic Hitting-Time Moment Approach to Seed-set Expansion over a Graph}

\author{
\name{A. Foss\textsuperscript{a,*}\thanks{afoss@sandia.gov} and R.B. Lehoucq\textsuperscript{a,*}\thanks{rblehou@sandia.gov} and Z.W. Stuart\textsuperscript{b}\thanks{zwstuart@unm.edu}  and J.D.Tucker\textsuperscript{a}\thanks{jdtuck@sandia.gov} and J. W. Berry\textsuperscript{a}\thanks{jberry@sandia.gov}}
\affil{\textsuperscript{a}Sandia National Laboratories}
\affil{\textsuperscript{b}Department of Mathematics and Statistics, The University of New Mexico}
\affil{\textsuperscript{*}These authors contributed equally to this work}
}

\maketitle

\begin{abstract}
We introduce HITMIX, a new technique for network seed-set expansion, i.e., the problem of identifying a set of graph vertices related to a given seed-set of vertices.
We use the moments of the graph's hitting-time distribution to quantify the relationship of each non-seed vertex to the seed-set.
This involves a deterministic calculation for the hitting-time moments that is scalable in the number of graph edges
and so avoids directly sampling a Markov chain over the graph.
The moments are used to fit a mixture model to estimate the probability that each non-seed vertex should be grouped with the seed set.
This membership probability
enables us to sort the non-seeds and threshold in a
statistically-justified way.
To the best of our knowledge, HITMIX is the first full statistical model for seed-set expansion that can give vertex-level membership probabilities.
While HITMIX is a global method, its linear computation complexity in practice
enables computations on large graphs.
We have a high-performance implementation, and we present computational
results on stochastic blockmodels and a small-world network from the SNAP
repository.  The state of the art in this problem is a 
collection of recently developed local
methods, and we show that distinct advantages in solution quality are available
if our global method can be used.  In practice, we expect to be able to run HITMIX if the graph can be stored in 
memory.
\end{abstract}


\section{Introduction}
\label{sec:intro}

\begin{figure}
\centering
\includegraphics[width=\textwidth]{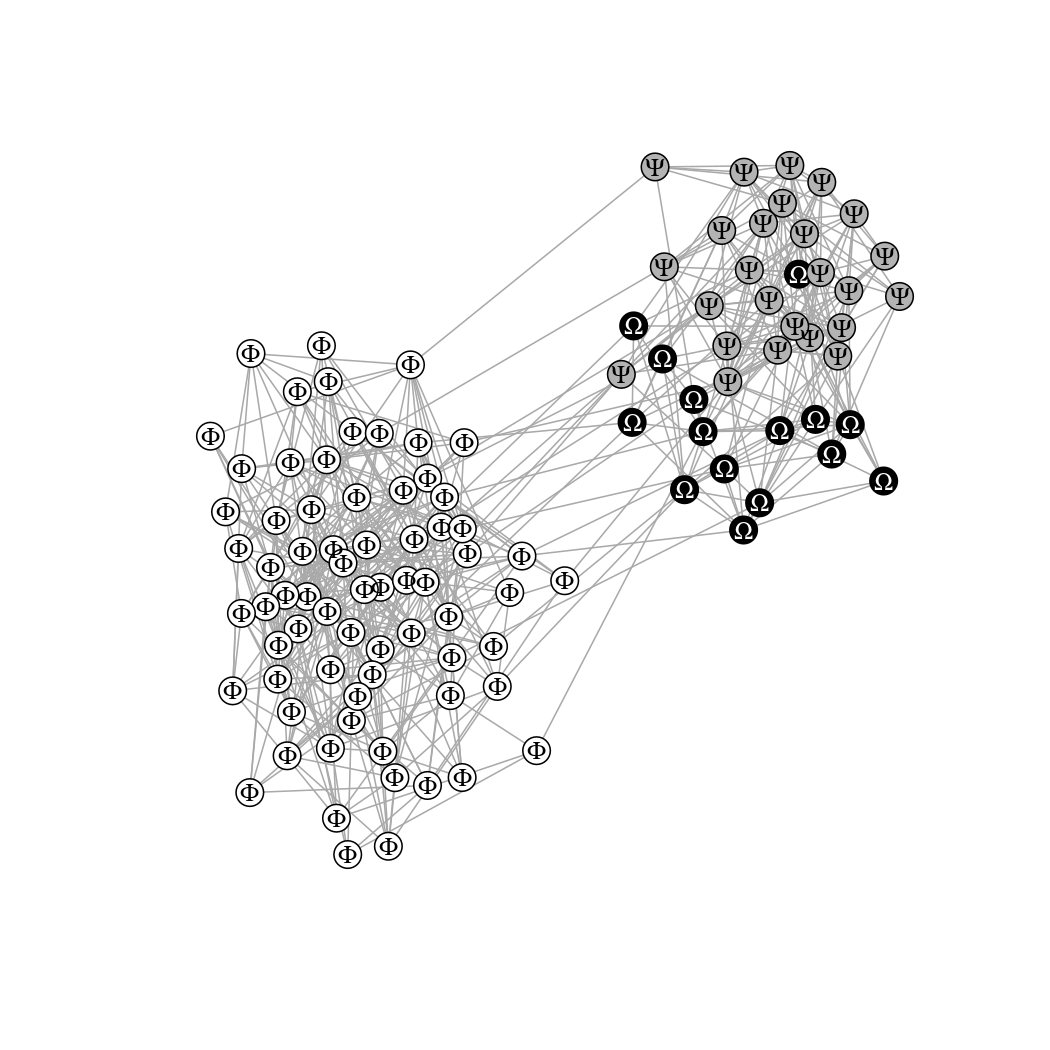}
\caption{
  \textit{Notation example}.
  This figure depicts the relationship between the seed-set vertices $\Omega$, the goal-set vertices $\Psi$ we seek to identify, and the non-goal set $\Phi$.
}
\label{fig:notation}
\end{figure}

We focus on partitioning a connected, undirected graph $\mathcal{G} = (\mathcal{V},\mathcal{E})$ with vertices $\mathcal{V}$ and edges  $\mathcal{E}$ with
respect to a seed-set of vertices $\hs \subset \mathcal{V}$. Specifically, we partition the non-seed vertices $\mathcal{V}\setminus \hs$ into two mutually exclusive sets $\Psi$ and $\Phi$ of vertices  so that $\mathcal{V} = \hs \cup \Psi \cup \Phi$.
The subset set $\Psi$ containing vertices with hitting-time probability distributions that favor reaching or hitting $\hs$ for a Markov chain evolving over $\Omega^c$.
We will refer to the vertices in $\Psi$ as the goal-set vertices, since we seek to identify $\Psi$ and distinguish its vertices from those in $\Phi$.
Figure~\ref{fig:notation} illustrates the relationship between $\Omega$, $\Psi$, and $\Phi$ for an example graph.

Our contribution is a novel method HITMIX, which provides the capability to estimate the set $\Psi$ given the seed-set $\Omega$. HITMIX is
\begin{itemize}
  \item scalable in the number of edges for the graph associated with $\Omega^c = \Psi \cup \Phi$,
  \item and provides a level of certainty for inclusion in $\Psi$ given a user specified threshold.
\end{itemize}
To the best of our knowledge, no published approach to seed-set expansion has
both of these properties.
We believe that this contribution will be of interest to the statistics,
computational mathematics, and data science communities.
Seed-set expansion has applicability across a wide variety of fields, such as identifying related brain regions in brain connectivity maps, identifying groups of related genes in gene networks, and identifying related entities in social networks.
Additionally, the probabilistic vertex-level group memberships provided by HITMIX allows a fine-grained inspection of membership strength, including the possibility that there is not an underlying block of related vertices.

\subsection{Related work}
\label{sec:related-work}
Our problem of determining a collection of vertices related to a given seed set (sometimes called \emph{network completion} or \emph{vertex nomination}) is a form of community detection and has been addressed by \cite{kloster:14,whgd:13} among others.
The general problem of finding communities in graphs is well-studied.
Fortunato's
canonical survey \cite{fortunato2010community} lists seven double-columned
pages of references and has itself been cited almost 10,000 times. However,
the discipline remains largely non-statistical in the sense that most
methods do not support statistical tests associating a given vertex
with a given community.  In other words, they do not provide an uncertainty
regarding community membership.  For example, in recent work by
Ha, et al.~\cite{hafm:20} the statistical guarantees provided are probabilistic
statements regarding the accuracy of a local clustering method on one
class of random graph models. However, their method does not quantify
vertex-specific uncertainties on cluster membership in real graphs.

Local graph clustering or partitioning, see e.g., \cite{ancl:06,spte:13,veldt16,veldt:19,wang:17} locates clusters given a seed node or set, with a running time that is independent of the size of the input graph. These methods attempt to identify partitions of $\mathcal{V} \setminus \hs$ that contain $\hs$ with small conductance by either sampling the graph or approximating the stationary distribution for a graph related to $\mathcal{G}$. This latter distribution is also popularly referred to as a localized or personalized PageRank vector where the seed vertices correspond to a teleportation vector; see \cite{glei:15} for a well-written review paper with numerous references and example applications.
The paper \cite{su17} employs random walks to estimate probabilities of vertex block membership.
Spectral approaches to the problem such as \cite{yoder2018vertexi} (where several related approaches are reviewed) have been applied effectively to
large stochastic block models.

We compared HITMIX to several standard local graph clustering algorithms: capacity releasing diffusion (CRD)~\cite{wang:17}, flow-seed (FS)~\cite{veldt:19}, Heat Kernel (HK)~\cite{kloster:14}, and simple local (SL)~\cite{veldt16}. CRD  is a \emph{push-relabel}\footnote{The numerical methods community refers to push algorithms as coordinate descent or relaxation methods.} algorithm that
uses techniques that are well-known from flow-based graph algorithms to identify a cluster.  FS and SL are flow based methods attempting to determine graph cuts  to minimize the conductance for vertices clustered with the seed vertices. HK approximates the localized PageRank vector via a truncated series that defines the heat-kernel for a graph.
We used Julia, Python, and MATLAB implementations provided by the authors of each method.
In Sections~\ref{sec:sbm} and \ref{sec:wikipedia} we provide preliminary information on the performance of HITMIX relative to some existing approaches.

\subsection{Our approach}
\label{sec:approach}

We use the moments of the hitting-time distribution to quantify the relationship of each non-seed vertex to the seed-set.
The mixture model is fit using a modified expectation-maximization (EM) algorithm that relies on the moment estimates to approximate vertex-specific hitting-time distributions. The mixture model enables us to rank-order the vertices in $\mathcal{V}\setminus \hs$ by certainty, or thresholded based on a desired certainty level. Hence given such a threshold, the set of vertices $\Psi$ related to $\Omega$ is determined.
The mixture model requires access to the vertex-level moment estimates, but not the graph itself, and uses pseud-random number generators to simulate samples from the hitting-time distributions.
This is more computationally efficient than sampling from the graph directly.
A discussion of the computational complexity of the EM algorithm is beyond the scope of this paper.
However, we note that the expected and observed performance of our EM algorithm is linear in the number of vertices.
The expected and observed runtimes for the conjugate gradient iteration method we use to obtain hitting-time moments are linear in the number of edges.

In contrast to other seed-set expansion techniques we tested, HITMIX was constructed to avoid difficult-to-set tuning parameters. We accomplish this by combining the conjugate gradient iteration and EM iterations, both well-known algorithms with easy-to-set parameters.
This is in contrast to the other seed-set expansion algorithms, which often require carefully choosing tuning parameters within an acceptable region, in which settings outside of this region fail to deliver useful results.
For example, diffusion-based methods (see, e.g., \cite{kloster:14} and references therein) generally involve parameters that control the influence of early vs late diffusion steps in selecting related vertices.
This decision is important, challenging, and is often dependent of the particulars of the graph in question.
We also remark, that these other methods tested are local, i.e. restricting access to a subgraph containing the seed vertices  whereas HITMIX is a global method because it requires hitting-time moments for all the non seed-set vertices. However, the conjugate gradient iteration leads to a high-performance implementation.

Although the hitting-time moments can be approximated via simulation, the computational cost is prohibitive and the simulation-based moments are inherently noisy. Instead, we exploit the fact that a deterministic estimator for the moments has a variational characterization in terms of a convex quadratic functional due to our assumption that the graph is connected and undirected. Hence the global minimum can be determined via the solution of a symmetric positive definite set of linear equations where the coefficient matrix is a submatrix of the graph Laplacian.
We solve this set of equations to an arbitrary level of precision using the conjugate gradient iteration, leading to an approach that is scalable in the number of graph edges associated with the subgraph of $\mathcal{G}$ associated with the vertices in $\mathcal{V} \setminus \hs$.

The remainder of this paper is organized as follows:
In Section \ref{sec:htprop}, we develop the hitting-time distribution and computation of the moments.
In Section \ref{sec:mixture} we describe the mixture model used the estimate vertex membership probabilities.
In Section \ref{sec:experiments} we report the results of applying the proposed approach to the stochastic block model and a real graph used to describe relationships among Wikipedia pages. Finally, we close with
  a brief summary and some ideas for future work in Section
  \ref{sec:conclusion}.


\section{A deterministic method for estimating hitting-time moments}
\label{sec:htprop}
The hitting-time to a collection of graph vertices $\hs $ is defined by the random variable
\[
T = T_{\hs} \coloneqq \min \{ j \geqslant 0 \colon X_j \in \hs \}\,,
\]
where $X_j$ is a random variable denoting the graph vertex occupied at the $j$-th step of a finite Markov chain.\footnote{We will drop the subscript $\hs$ for notational convenience.}
The probability mass function for the hitting-time to the set $\hs$ conditioned upon starting at vertex $i$ in $\nhs = \mathcal{V}\setminus \Omega$ is given by
\begin{align} \label{pmf-i}
   \prob(T_i = t) \coloneqq \prob(T = t \,|\, X_0 =i)\,.
\end{align}

Except for graphs with special structure, the $k$-th component of the mass function can only be determined via the matrix $(D^{-1}A)^k$, a prohibitive computation once the graph is not small. However, a deterministic approach for the hitting-time moments
\begin{align} \label{moments}
  \expect \, T^m_i
  & \coloneqq \expect(T^m = t \,|\, X_0 =i) = \sum_{t=1}^\infty T^m \, \prob( T^m_i = t )
\end{align}
is available. This avoids the prohibitively expensive approach of computing sample moments via simulation over the graph.

A standard derivation using first-step analysis\footnote{E.g., see \cite{norr:98}. The recursion we use is adapted from \cite{pala:09}.} demonstrates that the constrained linear set of equations
\begin{equation}\label{Eij}
  \left\{\begin{aligned}
            \expect T^m_i - \sum_{j\in \nhs} (e_i^{\top} D ^{-1}A \,e_j) \, \expect T^m_j & = 1 + \sum_{s=1}^{m-1} {m \choose s} \sum_{j\in \nhs} (e_i^\top D^{-1}A \,e_j)\expect  T^s_j  &  i \in \nhs \\
            \expect T^m_i  & = 0 &  i \in \hs
         \end{aligned}\right.
\end{equation}
provides a recursive formula for the $m$-th moment. Here, the notation $e_i^{\top} D ^{-1}A \,e_j$ denotes the element of the matrix $D ^{-1}A$ in row $i$ and column $j$.

If we eliminate the constraints in the previous equations, we may write a matrix equation for the relevant hitting-time moments for the vertices in $\nhs$ as
\begin{align} \label{ht-linear-system}
  \Pi_{\nhs}^\top \, (I-D ^{-1}A )\, \Pi_{\nhs}   \, \expect \, T^m  & =  e + \sum_{s=1}^{m-1} {m \choose s}  \Pi_{\nhs}^\top \,D ^{-1}A \, \Pi_{\nhs} \, \expect \, T^s \,,
\end{align}
where the vector $  \expect \, T ^m  $  contains the $\expect \,T^k_i  $ for $i \in \nhs$ and $e$ is the constant vector of all ones.
The prolongation  $\Pi_{\nhs} $ and restriction $\Pi_{\nhs}^\top$ matrices enable us to express the submatrix of $D^{-1}A$ corresponding to the vertices in $\nhs$ as $\Pi_{\nhs}^\top \,D ^{-1}A \, \Pi_{\nhs}$. The prolongation  $\Pi_{\nhs} $ is the submatrix that remains when removing the columns corresponding to the vertices in $\Omega$ from the identity matrix of order $n$.
Although the formula is for the (raw) moments, our paper focuses on the hitting-time means and variances $\mu_i  = \expect \,T_i$ and $\sigma_i^2    = \expect\, (T_i - \mu_i)^2$ respectively.

In contrast to estimating the moments via sampling the graph, the solution of the linear set of equations \eqref{ht-linear-system} provides an exact, deterministic estimate of the sample the moments.
Gaussian elimination is a standard method of solution for a linear set of equations. Unfortunately, the number of floating point operations for such an approach scales with the cubic power of the number of rows (and columns) of the coefficient matrix $\Pi_{\nhs}^\top \, (I-D ^{-1}A )\, \Pi_{\nhs}$ and neglects sparsity (that the vast majority of the elements are zero). Instead, we propose to use the conjugate gradient iteration~\cite{hest:52} as explained in Section \ref{sec:intro}. In order to do so, we rewrite \eqref{ht-linear-system}
 as
\begin{align} \label{ht-linear-system-spd}
  \Pi_{\nhs}^\top (I- \widehat{A})\Pi_{\nhs}\widetilde{\expect} \, T^m    & =  e + \Pi_{\nhs}^\top D ^{1/2}\Pi_{\nhs} \sum_{s=1}^{m-1} {m \choose s}  \Pi_{\nhs}^\top \widehat{A} \,\Pi_{\nhs}\, \widetilde{\expect} \, T^s \,,
  \intertext{where}
   \widetilde{\expect} \, T^m & = \Pi_{\nhs}^\top D ^{1/2}\Pi_{\nhs} \expect \, T^m \nonumber \\
   \widehat{A} &= D ^{-1/2}AD ^{-1/2} \nonumber
\end{align}
and we used the (well-known) identity
 $D^{1/2}(I-D ^{-1}A)D ^{-1/2}  = I-\widehat{A}.$

This identity demonstrates that the shifted transition matrix is related to the normalized graph Laplacian via a similarity transformation. Because of our assumption that the graph is connected and undirected, the coefficient matrix $\Pi_{\nhs}^\top \big(I-\widehat{A} ) \Pi_{\nhs} $ is symmetric and positive definite so that we may use the conjugate gradient iteration to approximate the solution $\widetilde{\expect} \, T^m$. We remark that the needed moments $\expect \, T^m$ are easily recovered  since $\expect \, T^m = \Pi_{\nhs}^\top D ^{-1/2}\Pi_{\nhs}\widetilde{\expect} \, T^m$. An important, practical, aspect of the conjugate gradient iteration is that only the application of the coefficient matrix in \eqref{ht-linear-system-spd} to a vector is necessary. The number of floating point operations for each of these matrix-vector applications is roughly proportional to the twice the number of edges present. It is this property and simple implementation, along with its well-understood mathematical properties, including its variational characterization
\begin{align}\label{var-char}
  \widetilde{\expect} \, T^m & =\argmin_{x} \frac{1}{2} x^\top H x - x^\top b
\end{align}
where $H$ and $b$ are the coefficient matrix and the righthand side vector of \eqref{ht-linear-system-spd},
which makes the conjugate gradient iteration an extremely powerful method for approximating the solution of \eqref{ht-linear-system-spd}.

\section{Mixture Model}
\label{sec:mixture}

Our goal is to estimate a set $\ns$ of vertices that are closely related to the seed set $\hs$.
As we explained in Section \ref{sec:intro}, we use a mixture model to provide an estimated probability of membership $\widehat{\prob}(v_i \in \ns)$ for the vertices $v_i \in \nhs$.
These estimated probabilities reflect the level of certainty with which the model has allocated a vertex to $\ns \subseteq \Omega^c$; this allows vertices to be rank-ordered by certainty, or thresholded based on a desired certainty level. Recall from the discussion in Section \ref{sec:intro} that this allows us to partition the vertices $\Omega^c = \Psi \cup \Phi$ into two mutually exclusive sets.

The inputs to the mixture model are samples of the hitting time probability mass function for each vertex in $\nhs$.
As explained following \eqref{pmf-i}, these probability mass functions are not generally available and so must be estimated.
Our approach is to use a method of moments estimator for a chosen distribution family.
Although our model could be modified to use any distribution and any number of moments, the results presented in this paper use the first two moments (mean, variance) to fit the lognormal distribution.
We found the lognormal distribution works well in practice, and offers computationally fast closed-form estimators.
The choice of distributions was also influenced by results of Lau and Szeto~\cite{lau10}, who show that after a sufficient relaxation time, the hitting-time distribution is proportional to $\alpha \exp(-\beta t)$.
This suggests that hitting-time distributions should be modeled by distributions with exponentially decaying tails, including (but not limited to) the lognormal distribution.

Recall that $T_i$ denotes a random variable describing the hitting time of vertex $i$ in $\nhs$ with respect to $\hs$, see \eqref{pmf-i}.
The mixture model summarizes the hitting time distributions for vertices in $\ns$  using an aggregate probability mass function (PMF) $f(t; \vt{1})$, while vertices in $\Phi$ are summarized with distinct distributions from the same family, $f(t; \vt{k})$ for $2 \leq k \leq g$:

\begin{align*}
  \prob(T_i = t | v_i \in \Psi) &\coloneqq f(t; \; \vt{1})\\
  \prob(T_i = t | v_i \in \Phi) &\coloneqq f(t; \; \vt{k}), \quad k \in \{2, 3, ..., g\}
\end{align*}

The number of clusters $g$ may be specified manually based on subject-matter expertise, or selected automatically.
Selecting $g > 2$ offers the potential to model greater heterogeneity and/or distinct subpopulations in the data.
For automatic selection of $g$, we use the Bayesian Information Criterion (BIC; \cite{schwarz}).
BIC is a widely-used metric that encourages parsimonious models by rewarding goodness-of-fit, but penalizing a larger number of parameters.
We fit the mixture model for a plausible range of $g$, and select the model that optimizes the BIC.

The components $f(t; \vt{k})$ are currently selected to be lognormal.
Thus, we model $T_i$ using a standard mixture distribution with PMF
\begin{equation}
   \mathbb{P}(T_i = t) \coloneqq \sum_{k=1}^g \pi_k f(t; \; \vt{k}),
\label{eq:mixMod}
\end{equation}
\noindent where $\pi_k$ is the prior probability of class $k$.
Conditional on known class membership, this simplifies to $ \mathbb{P}(T_i = t | v_i \in \text{class} \, k) \coloneqq f(t; \; \vt{k}). $


For each $T_i$, draw $m$ pseudo-random samples from the fitted distributions.
In contrast to directly sampling hitting-times from the graph, this sampling procedure does not require access to the graph, leading to substantially faster computation.
The likelihood of the unknown model parameters with regard to the data sample is

\begin{align*}
  L(\Theta; \cod, \curv) &= \prod_{i \in \nhs} \prod_{j=1}^m \sum_{k=1}^g Z_{ik} \mathbb{P}(T_i = t_{ij}; \vt{k}), \\
  &= \prod_{i \in \nhs} \prod_{j=1}^m \sum_{k=1}^g Z_{ik} f(t_{ij}; \vt{k}),
\end{align*}
\noindent where $\cod$ is the collection of the observed data $t_{ij} \; \forall i \; \forall j$, $\curv$ is the collection of the unobserved random variables $Z_{ik} \; \forall i \; \forall k$, $\Theta$ is the collection of all unknown parameters $\vt{k}$, $\pi_k \; \forall k$, and where $Z_{ik}$ is a random variable equal to 1 if $T_i$ has PMF $f(\vt{k})$ and equal to 0 otherwise. Thus, the complete data log-likelihood can be written as

$$ l(\Theta; \cod, \curv) = \sum_{i \in \nhs} \sum_{j=1}^m \sum_{k=1}^g Z_{ik} \log f(t_{ij}; \vt{k}). $$

The classical way to generate max likelihood estimators in the presence of unobserved information is the EM algorithm \cite{dempster77}, which proceeds to maximize the expected log-likelihood in two iterative, alternating steps. The E step involves constructing the $Q$ function as
$$ Q(\Theta | \hat{\Theta}^{(w)}) \coloneqq \underset{\curv | \cod, \Theta = \hat{\Theta}^{(w)}}{\expect}[l(\Theta; \cod, \curv)], $$
\noindent where $w$ indexes EM iteration. The M step involves maximizing the $Q$ function to obtain the next estimate of $\Theta$:
$$ \hat{\Theta}^{(w+1)} \coloneqq \underset{\Theta}{\arg\max} \, Q(\Theta|\hat{\Theta}^{(w)}). $$

The E step in this case amounts to finding the expected value of the $Z_{ik}$.
Applying Bayes' rule, we obtain
\begin{align*}
  \expect[Z_{ik} \, | \, \cod, \hat{\Theta}^{(w)}] &= \mathbb{P}(T_i \in \text{class} \; k \, | \, \cod, \Theta = \hat{\Theta}^{(w)}) \\
  &= \mathbb{P}(t_{i1}, t_{i2}, ..., t_{im} \;  \in \text{class} \; k\, | \, \cod, \Theta = \hat{\Theta}^{(w)}) \\
  &= \prod_{j=1}^m \mathbb{P}(t_{ij} \;  \in \text{class} \; k\, | \, \cod, \Theta = \hat{\Theta}^{(w)}) \\
  &= \prod_{j=1}^m \frac{\hat{\pi}_k^{(w)} f(t_{ij} \, | \, \hvt{k}^{(w)})}{\sum_{\ell=1}^g \hat{\pi}_\ell^{(w)} f(t_{ij} \, | \, \hvt{k}^{(w)})} \\
\end{align*}

The M step in this case amounts to a standard weighted maximum-likelihood estimator of the parameter vector $\vt{k}$ for each component, with score functions
$$\frac{\partial}{\partial \vt{\ell}}Q(\Theta|\hat{\Theta}^{(w)}) = \sum_{i \in \nhs} \sum_{j=1}^m \expect[Z_{ik} \, | \, \cod, \hat{\Theta}^{(w)}] \frac{\partial}{\partial \vt{\ell}} \mathbb{P}(T_i = t_{ij}; \vt{\ell}). $$
Each $\pi_\ell$ must also be estimated in the M step.
There are multiple strategies available depending upon the prior information available.
In the absence of strong prior information, we can use
$$\hat{\pi}_k^{(w+1)} = \sum_{i \in \nhs} \expect[Z_{ik} \, | \, \cod, \hat{\Theta}^{(w)}].$$

The EM algorithm involves alternating between the E and M steps, for some initial estimator $\Theta^{(0)}$, until a local optimum is reached.

\subsection{HITMIX}

The discussion in Section \ref{sec:htprop} and in this section are summarized in the HITMIX algorithm.
HITMIX estimates a set $\ns \subset \mathcal{V}\setminus \Omega$ of vertices that are related to the seed set $\hs$ and consists of the following four steps:
\begin{enumerate}
  \item compute the hitting-time means and variances for the vertices $\mathcal{V}\setminus \Omega$ via the solution of the linear set of equations \eqref{ht-linear-system},
  \item use a method of moment estimator to model the hitting-time distributions for the vertices in $\mathcal{V}\setminus \Omega$ given the means and variances,
  \item use a lognormal mixture model to estimate the membership probabilities,
  \item given a threshold $0\leqslant \tau \leqslant 1$ on the membership probabilities, set $\ns$ to the vertices larger than the threshold.
\end{enumerate}

The solution of the linear set of equations in step (1) is accomplished by reformulating \eqref{ht-linear-system} as a symmetric positive definite system \eqref{ht-linear-system-spd}, see Section~\ref{sec:htprop}.  We then approximate the means and variances by using the conjugate gradient iteration, which requires us to specify a tolerance for terminating the iteration and we initialize the  iteration with a random starting vector. We remark that only this step requires access to the graph.

The method of moments (MOM) estimator in step (2) is accomplished separately for each vertex.
Given central moment estimates $m_1$ and $m_2$, the MOM estimators for the lognormal distribution are
\begin{align*}
  \hat{\sigma}^2_{\text{mom}}  \coloneqq \log \left( \frac{m_2}{m_1^2} + 1 \right) \,, \text{ and }
  \hat{\mu}_{\text{mom}}  \coloneqq \log \left( m_1 \right) - \frac{\hat{\sigma}_{\text{mom}}^2}{ 2} \, .
\end{align*}

Step 3 involves fitting model~\ref{eq:mixMod} using $k=2$ and the lognormal distribution.
Step 4 involves assigning hard classifications with $\tau = 0.5$.

\section{HITMIX Experiments}
\label{sec:experiments}


We begin with a set of experiments on synthetic data.
We focus on generating data with simple and known structure to facilitate straightforward comparisons between algorithms.
Performing well on such data should not be taken as evidence that an algorithm is generally effective, however, performing poorly on such data is of concern.
We then analyze a large real-world dataset that is more representative of problems of interest.

\subsection{Stochastic Block Model}
\label{sec:sbm}

In order to investigate the properties of the proposed method, we conducted a set of Monte Carlo simulation studies using the stochastic block model (SBM)~\cite{hobl:83}.
The Monte Carlo parameter settings are described in Table~\ref{tab:sbmSettings}.
In each SBM simulation, two or more blocks were specified, and the first was treated as the goal block.
A subset of the vertices in the goal block were randomly sampled and treated as the hitting set.
We ran our algorithm, and compared the identified neighbors to the true membership in the goal block using the Adjusted Rand Index (ARI; \cite{hubert85}) and $F_1$ scores.
For the calculation of $F_1$ scores, the goal block was identified as the block with the smallest mean hitting times.
In all SBM simulations, we drew 500 Monte Carlo samples for each condition. 

\begin{table}
\begin{center}
\begin{tabular}{llll}
\toprule
Parameter & Simulation 1 & Simulation 2 & Simulation 3 \\
\midrule
Num. Blocks ($b$) & 2,3, ..., 10 & 2 & 2 \\
Block Size &   200         & 100            & 100 \\
Hitting Set Size & 20 & 10 & 1, 5, 10, 25, 50 \\
$p_{in}$   & 0.15 & 0.05, 0.06, ..., 0.20 & 0.15 \\
$p_{out}$  & $0.05 / (b-1)$ & 0.05 & 0.05 \\
MC Samples & 500 & 500 & 500 \\
\bottomrule
\end{tabular}
\end{center}
\caption{Monte Carlo Parameter Settings for SBM Simulations}
\label{tab:sbmSettings}
\end{table}

\begin{figure}
  \centering
  \begin{subfigure}[t]{.48\linewidth}
    \centering\includegraphics[width=\textwidth]{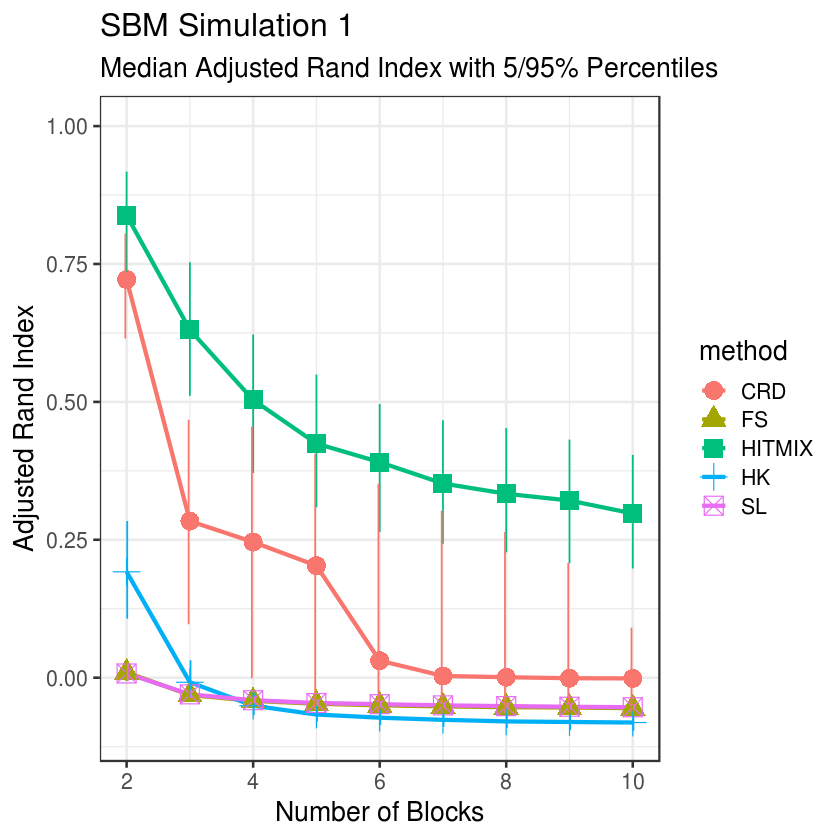}
    \caption{ARI}\label{fig:sbmBlockAri}
  \end{subfigure}
  \begin{subfigure}[t]{.48\linewidth}
    \centering\includegraphics[width=\textwidth]{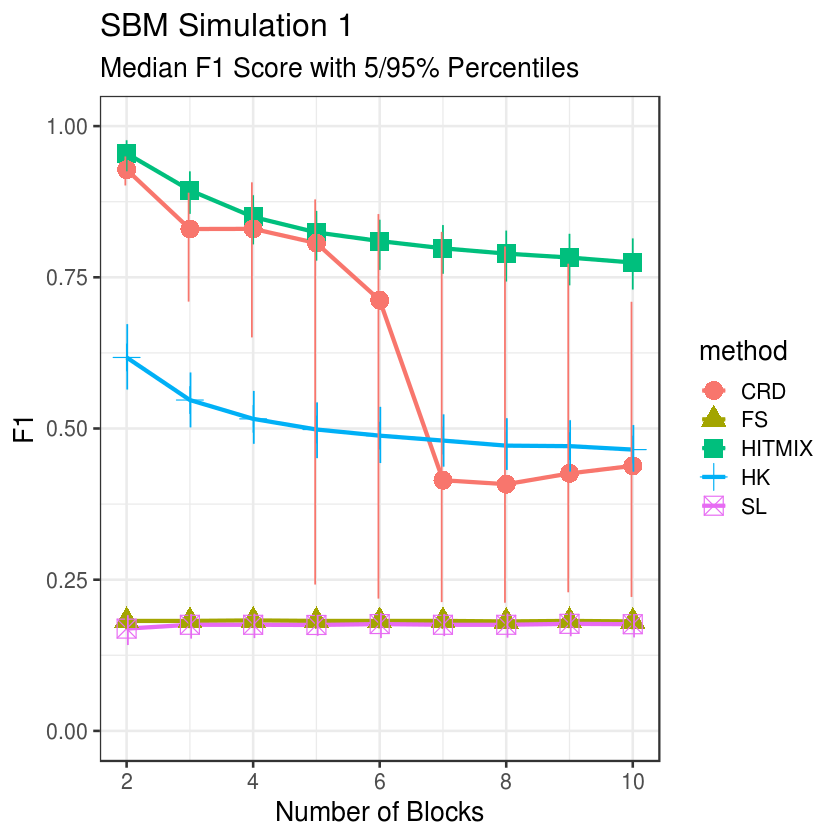}
    \caption{$F_1$}\label{fig:sbmBlockF1}
  \end{subfigure}
  \caption{ARI and $F_1$ scores by number of blocks in SBM Simulation 1. Parenthetical values are 5 and 95\% percentiles. Percentiles are taken across Monte Carlo runs. 500 Monte Carlo samples were taken per setting of number of blocks.}
  \label{fig:sbmBlock}
\end{figure}

In SBM Simulation 1, the number of blocks was varied from 2--10.
Block size was fixed at 200, and the probability of in-block edges was fixed at 0.15.
The probability of out-block edges was set to $0.05/(b-1)$ in order to hold the expected number of out-block edges constant ($b$ denotes the number of blocks).
As shown in Table~\ref{tab:sbmBlock} and Figures~\ref{fig:sbmBlockAri} and \ref{fig:sbmBlockF1}, performance decreased as the number of blocks increased, but remained well above chance level (an ARI of 0 corresponds to chance-level partitions).

\begin{figure}
  \centering
  \begin{subfigure}[t]{.48\linewidth}
    \centering\includegraphics[width=\textwidth]{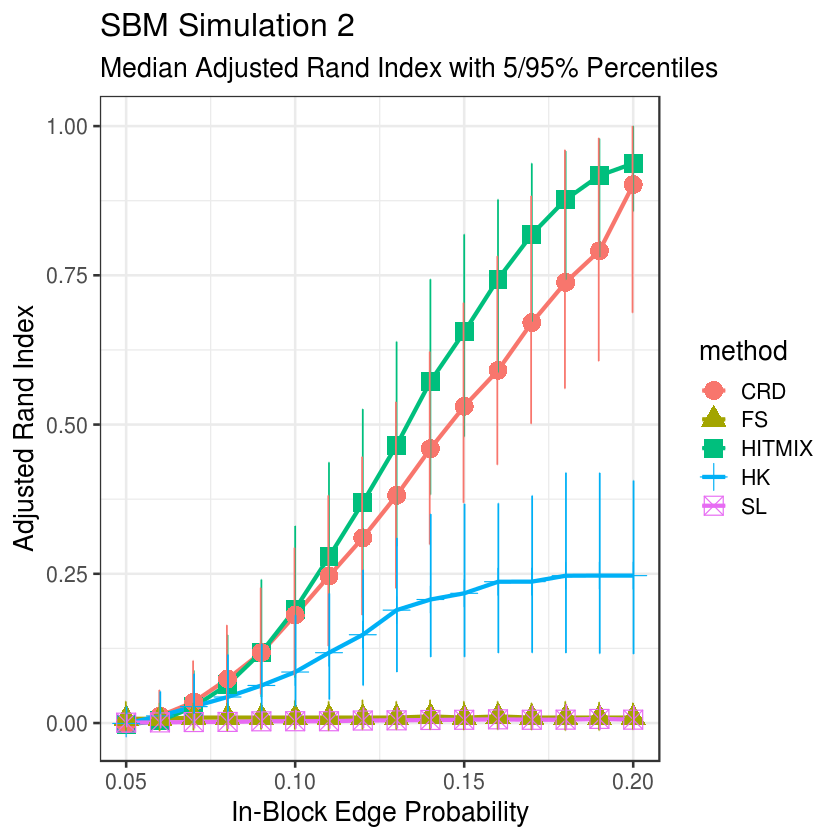}
    \caption{ARI}\label{fig:sbmPinAri}
  \end{subfigure}
  \begin{subfigure}[t]{.48\linewidth}
    \centering\includegraphics[width=\textwidth]{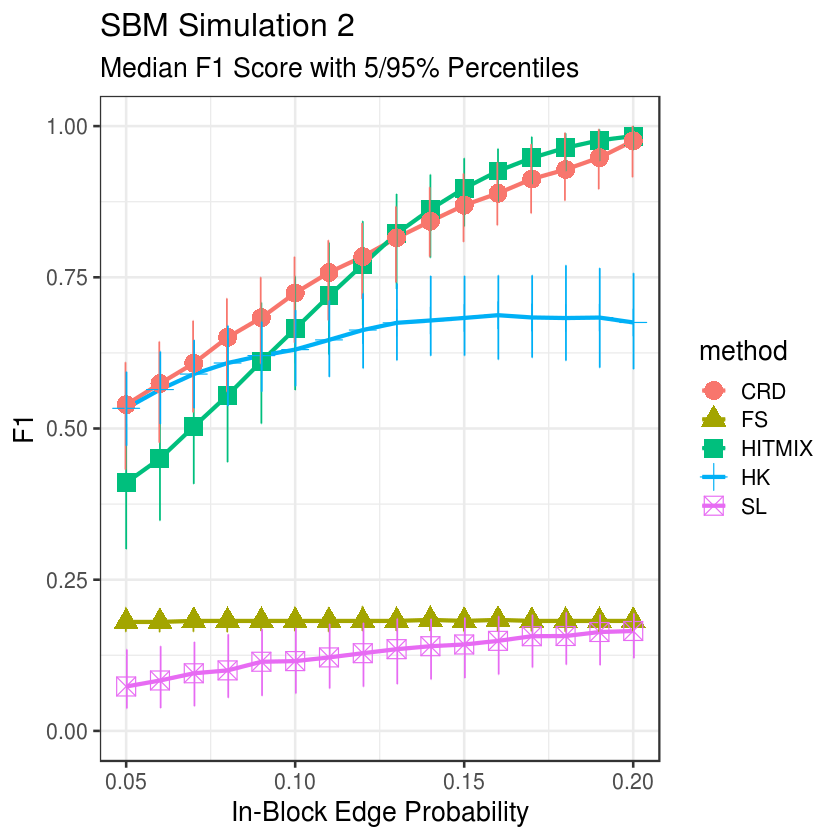}
    \caption{$F_1$}\label{fig:sbmPinF1}
  \end{subfigure}
  \caption{ARI and $F_1$ scores by $p_{in}$ in SBM Simulation 2. Parenthetical values are 5 and 95\% percentiles. Percentiles are taken across Monte Carlo runs. 500 Monte Carlo samples were taken per setting of $p_{in}$.}
  \label{fig:sbmPin}
\end{figure}

In SBM Simulation 2, two-block SBM graphs were sampled.
The probability of an out-block edge was fixed at 0.05, while the probability of an in-block edge $p_{in}$ was varied from 0.05 to 0.2 in increments of 0.01.
Block size was fixed at 100.
As shown in Table~\ref{tab:sbmHit} and Figures~\ref{fig:sbmPinAri} and \ref{fig:sbmPinF1}, performance varied from nearly perfect for $p_{in} = 0.2$, and decreased to chance-level performance at around $p_{in} = 0.06$.

\begin{figure}
  \centering
  \begin{subfigure}[t]{.48\linewidth}
    \centering\includegraphics[width=\textwidth]{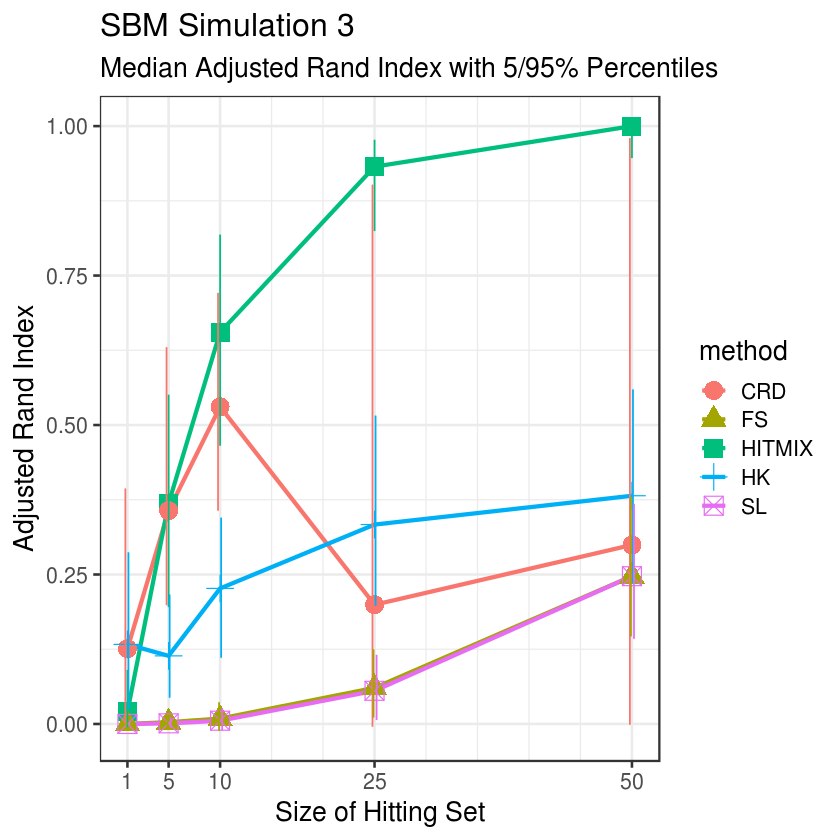}
    \caption{ARI}\label{fig:sbmHitAri}
  \end{subfigure}
  \begin{subfigure}[t]{.48\linewidth}
    \centering\includegraphics[width=\textwidth]{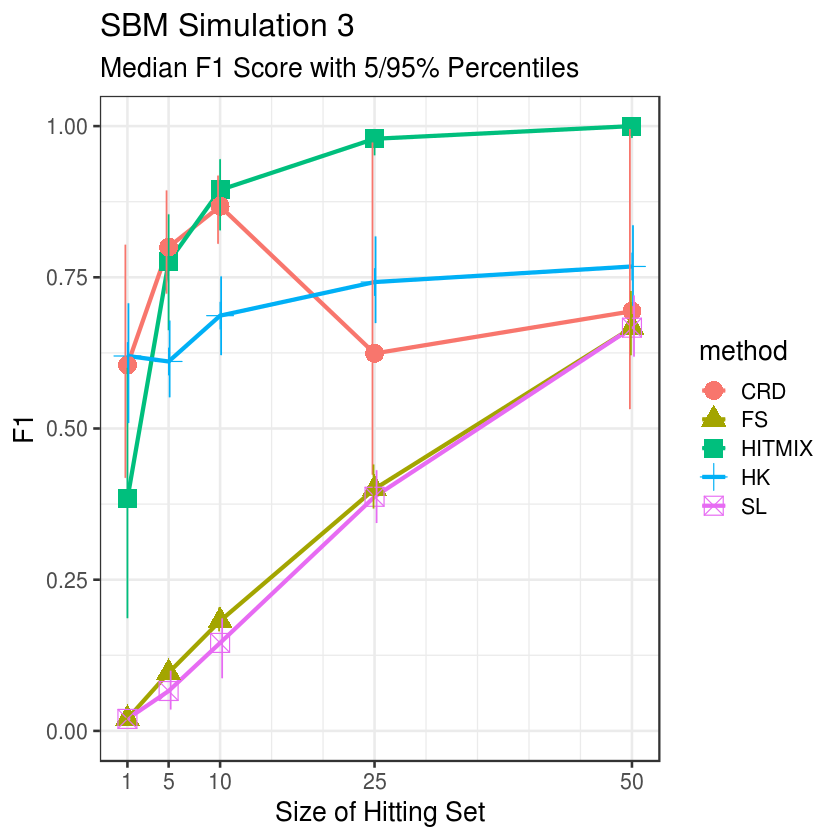}
    \caption{$F_1$}\label{fig:sbmHitF1}
  \end{subfigure}
  \caption{ARI and $F_1$ scores by hitting set size in SBM Simulation 3. Parenthetical values are 5 and 95\% percentiles. Percentiles are taken across Monte Carlo runs. 500 Monte Carlo samples were taken per setting of hitting set size.}
  \label{fig:sbmHit}
\end{figure}

In SBM Simulation 3, two-block SBM graphs were sampled, with 100 vertices per block.
The probability of in- and out-block edges were fixed at 0.15 and 0.05, respectively.
In order to investigate the relationship between hitting-set size and neighbor detection performance, we varied the size of the hitting-set from 1--50.
As shown in Table~\ref{tab:sbmHit} and Figures~\ref{fig:sbmHitAri} and \ref{fig:sbmHitF1}, performance was excellent for hitting-sets comprising 50\% and 25\% of the goal block, and then began to decline sharply until reaching nearly chance-level performance for a hitting-set of one single vertex.

\section{Wikipedia Network of Top Categories}
\label{sec:wikipedia}

To demonstrate our method on a large real-world graph, we utilized the Wikipedia topcats dataset from the Stanford Network Analysis Project (SNAP)~\cite{wikiSnap}.
The network is an undirected graph comprised of Wikipedia hyperlinks collected on September 2011.
Vertices represent articles, edges represent hyperlinks between pages, and communities represent article categories.
The graph is a subset of Wikipedia: the data curators first took the largest strongly connected component of the graph, and then restricted to article categories with at least 100 pages.
The dataset contains about 1.8 million vertices, 28.5 million edges, and 17 thousand article categories.
Article categories are not mutually exclusive, and span a wide range of topics, e.g. algebraic geometry, state highways in Missouri, asteroids named for people, Florida State University faculty, living people, etc.
The diameter of the graph (the longest shortest path) is 9, and the 90th percentile shortest distance between all node pairs is 3.8 (i.e. 90th percentile effective diameter) \cite{wikiSnap}.

We constructed a goal set (i.e., vertices to be detected) by merging the categories ``American film actors'' and ``American films,'' representing 29,240 vertices.
This set shared about 1.16 million unique edges with the rest of the graph, and had a volume of about 195 thousand (edges between members of the set).
The ratio of these two quantities (conductance \cite{veldt16}) was approximately 6.
We randomly selected 50\% of these vertices as seeds, with the remaining nodes reserved as vertices to be detected.
The conductance of the seed set was approximately 16.

We analyzed the resulting graph using HITMIX using a lognormal mixture model.
\af{Need basic hardware details for HPPDA.}
Using all 20 cores of a shared-memory machine with Intel(R) Xeon(R) CPU E5-2698 v4 \@ 2.20GHz processors, calculating the moments with a high-performance C++ code took about 9.6 seconds; most of this time was file I/O and constructing matrices in memory.
Solving the linear systems to obtain the first and second moments took a total of 0.65 seconds.
We specified two through five clusters in the mixture model, and selected the best model using the BIC as described in Section~\ref{sec:mixture}.
\af{Need basic hardware details for Sisu.}
Using 4 cores of a shared-memory machine with 80 Intel(R) Xeon(R) CPU E7-8891 v3 \@ 2.80GHz processors, fitting the mixture model with $g = 2, 3, 4,$ and $5$ using R took 2.1, 3.0, 2.8, and 3.2 hours, respectively.
Since the EM runs for each setting of $g$ do not depend on each other, they can easily be run in parallel.
We thresholded all posterior group probabilities at 0.5 to obtain hard group memberships.
We also analyzed the graph with other methods described in Section~\ref{sec:intro}.

\begin{figure}
\centering
\includegraphics[width=\textwidth]{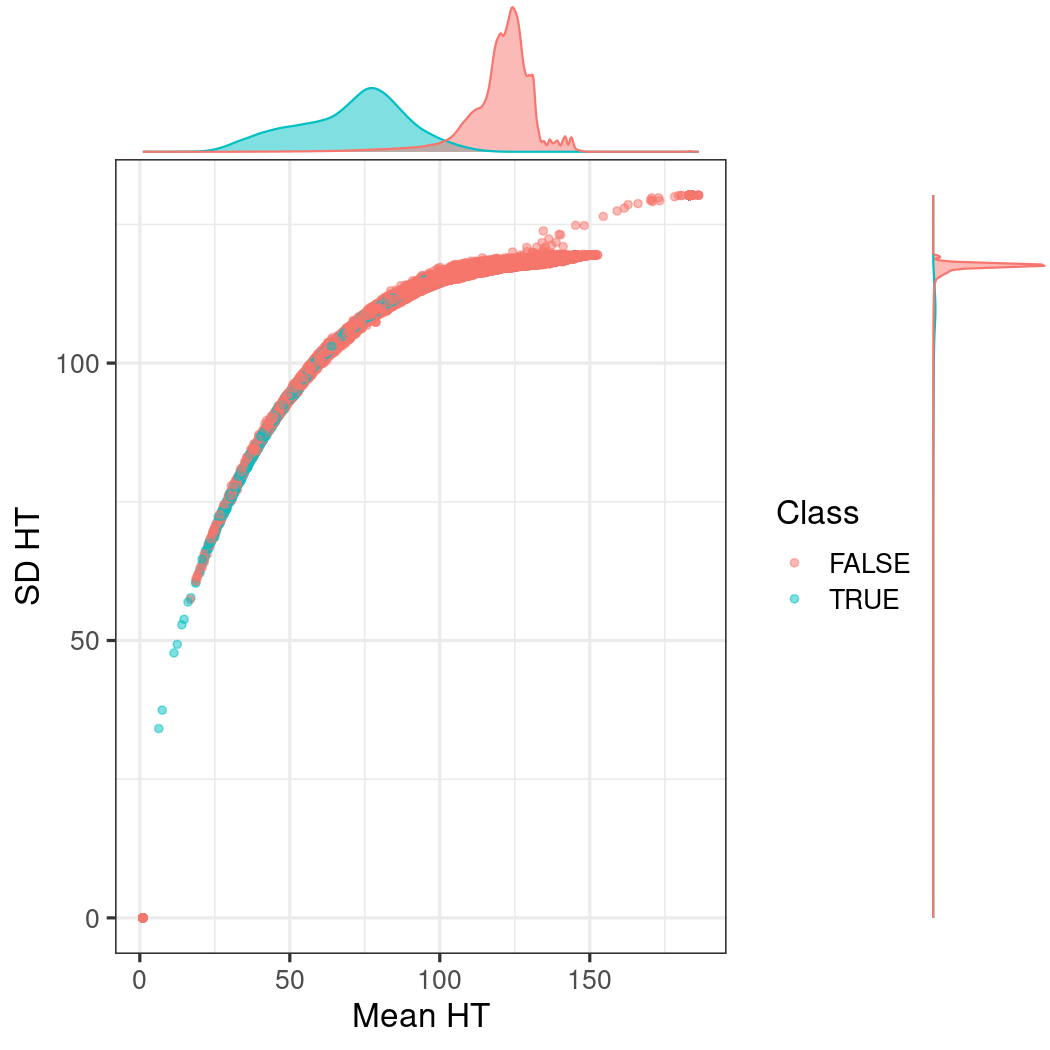}
\caption{
  \textit{Wikipedia graph moments}. This scatterplot shows the calculated mean and standard deviation hitting times for the entire wikipedia graph. Each point represents one article, colored blue if a member of the goal set and red otherwise.
The top density plot shows the distribution of the mean hitting times, normalized to equal area within goal and non-goal vertices.
The side density plot shows the distribution of the standard deviations, with area under the curve proportional to relative counts of goal and non-goal vertices.
}
\label{fig:wikiMoments}
\end{figure}

Figure~\ref{fig:wikiMoments} shows the mean and standard deviations of each vertex.
There is a remarkably close curvilinear relationship between the two moments with some notable exceptions.
The small curve in the top right of the plot corresponds to articles in the topic buprestoidea (a super-family of metallic wood-boring beetles).
The point mass near (1,0) appears to represent articles that link exclusively to the seed set (after the preprocessing step described above), based on a manual inspection of the graph data and the September 2011 version of Wikipedia using the ``View History'' feature of Wikipedia \cite{wikiHistory}.
Recall that in the downloaded data set, articles from categories with less than 100 pages were removed, along with any associated links.
For example:
\begin{itemize}

\item The 2011 article on ``Ella Baff''%
\footnote{
  Ella Baff (July 2011 revision),
  \url{https://en.wikipedia.org/w/index.php?title=Ella_Baff&oldid=437294603},
  accessed September 14, 2020
}
(mean = 1.0, SD = 0.00044), a distinguished arts consultant, had a single surviving edge to ``Regina Baff,'' an American film actress%
\footnote{
  Regina Baff,
  \url{https://en.wikipedia.org/wiki/Regina_Baff},
  accessed September 14, 2020
}%
.

\item The 2011 article on ``Cartilage-hair hypoplasia''%
 \footnote{
  Cartilage-hair hypoplasia (August 2011 version),
  \url{https://en.wikipedia.org/w/index.php?title=Cartilage\%E2\%80\%93hair_hypoplasia&oldid=443626655},
  accessed September 14, 2020,
 }
 (mean = 1.0, SD = 0.00025), a rare genetic disorder, had two surviving edges: one to ``Billy Barty,'' an American film actor and founder of the Little People of America organization%
 \footnote{
  Billy Barty,
  \url{https://en.wikipedia.org/wiki/Billy_Barty},
  accessed September 14, 2020,
 }
 and the second to ``Verne Troyer,'' an American film actor most famous for his role of Mini-Me in the Austin Powers film series%
 \footnote{
  Verne Troyer,
  \url{https://en.wikipedia.org/wiki/Verne_Troyer},
  accessed September 14, 2020,
 }%
 .

\item The 2011 article on ``Love on Toast''%
 \footnote{
  Love on Toast (August 2011 version),
  \url{https://en.wikipedia.org/w/index.php?title=Love_on_Toast&oldid=444068983},
  accessed September 14, 2020
 }
 (mean = 1.0, SD = 0.00040), an American comedy film, had two surviving edges: one to ``John Payne,'' an American film actor%
 \footnote{
  John Payne (actor),
  \url{https://en.wikipedia.org/wiki/John_Payne_(actor)},
  accessed September 14, 2020,
 }%
 , and one to ``Isabel Jewell,'' an American film actress%
 \footnote{
  Isabel Jewell,
  \url{https://en.wikipedia.org/wiki/Isabel_Jewell},
  accessed September 14, 2020,
 }%
 . Although both links are included in the dataset, Isabel Jewell does not appear on the archived page in Wikipedia.

\end{itemize}

The results of HITMIX compared to other techniques is shown in Table~\ref{tab:wikiF1}.
HITMIX for clusters 2--5 outperforms other methods on the basis of $F_1$ score.
We note that all differences by method in precision and recall are significant based on McNemar's $\chi^2$ test at the 0.001 level with continuity correction, although this is not the same as practically different.
The 5-cluster HITMIX solution was selected on the basis of the BIC.
A higher number of HITMIX clusters yields higher precision (at the cost of recall), since more clusters allows a more flexible model on the non-goal vertices, resulting in fewer nominations of candidate vertices.
The Heat Kernel method and SimpleLocal did not perform well, presumably since the conductance of the goal vertices was quite high as discussed above.
For some methods, it was unclear how to set tuning parameters; we sought to mitigate this challenge by using a range of parameters and reporting the best results.
\begin{table}[ht]
\centering
\begin{tabular}{lrrr}
  \toprule
 Method & $F_1$ & Precision & Recall \\ 
  \midrule
 HITMIX, 2 Clusters & 0.21 & 0.12 & 0.96 \\ 
  HITMIX, 3 Clusters & 0.34 & 0.21 & 0.85 \\ 
  HITMIX, 4 Clusters & 0.39 & 0.28 & 0.65 \\ 
  HITMIX, 5 Clusters & 0.37 & 0.34 & 0.39 \\ 
  CRD & 0.19 & 0.11 & 0.80 \\ 
  HK &  -- & 0.00 & 0.00 \\ 
  SL &  -- &  -- & 0.00 \\ 
   \bottomrule
\end{tabular}
\caption{Results of wikipedia-topcats analysis. HITMIX results for 2 -- 5 clusters are shown here. Note that HITMIX selected the 5 cluster solution on the basis of the BIC.}
\label{tab:wikiF1}
\end{table}

Across all methods, the number of false positives was quite high (as reflected by low precision scores, all below 0.34).
A follow-up analysis revealed that the largest amount of false positives were from categories that were sensible given the goal set (recall that the goal set is a union of American film actors and American films).
For example, in the HITMIX 2-cluster solution, we inspected the first category listed for each false positive article (recall that each page can have an arbitrary number of categories).
Except for the first category (``living people,'' 4,213 articles), nearly all of the top 50 categories tabulated were closely related to the goal set. In descending order, the next categories were: ``American film directors'' (1,576), ``American actors'' (1,263), ``2000s American television series'' (1,146), ``2000s drama films'' (979), ``American screenwriters'' (797).

\section{Conclusion and Future Work}
\label{sec:conclusion}

In this paper we present the novel algorithm HITMIX for seed-set expansion based on hitting-time moments.
We introduced a deterministic method for calculating hitting-time moments via the solution of the linear set of equations that scales in the number of edges, and avoids costly Monte Carlo sampling.
We then introduce a mixture model for partitioning vertices into two subgraphs: those that are related to the seed set and those that are unrelated given a prescribed threshold.

In contrast to other seed-set expansion techniques we tested, HITMIX was constructed to avoid difficult-to-set tuning parameters. We accomplish this by combining the conjugate gradient iteration and EM iterations, both well-known algorithms with easy to set parameters.
In contrast, other methods often require tuning parameters within an acceptable zone outside of which the results are unpredictable.
In contrast, for HITMIX, increasing the number of EM iterations and of the pseudorandom samples, and decreasing the conjugate gradient iteration relative residual tolerance improves the analysis. Specification of these parameters is largely dictated by the available computational resources. The one exception, the number of clusters $g$, can be auto-selected using BIC.
%
Often, if the goal vertices are expected to be well-separated from the rest of the graph, setting $g=2$ will suffice, as the hitting time moments will fall into two separable groups.
We also remark, that these other methods tested are local, i.e. restricting access to a subgraph containing the seed vertices  whereas HITMIX is a global method because it requires hitting-time moments for all the non seed-set vertices. However, the conjugate gradient iteration leads to a high-performance implementation.

Across our synthetic and real-world results, we see that alternative methods perform well in certain cases but not others.
HITMIX is the only method that delivers consistently good results across a wide range of conditions.
In particular, HITMIX is able to perform well for the identification of well-separated subgraphs, as in the SBM simulations (Section~\ref{sec:sbm}), and also perform well for subgraphs that share a substantial number of edges with adjacent vertices, as in the Wikipedia analysis (Section~\ref{sec:wikipedia}).

The current mixture model requires estimates of the first two moments of each vertex but can  be adapted to include higher-order moments.
As discussed in Section~\ref{sec:htprop}, the estimator can recursively estimate an arbitrary number of moments.
These higher-order moments can be used to improve the method of moments estimation scheme described in Section~\ref{sec:mixture}.
Further study is required to identify the conditions in which using higher-order moments could benefit the current framework.

Because of the variational characterization \eqref{var-char} for the hitting-time moments, the conjugate gradient iteration may be replaced with other optimization approaches. For instance, a coordinate descent method can be used so restricting access to the graph. Moreover, although our focus in this paper is on undirected graphs, hitting-time moments are well-defined for directed graphs however a variational characterization is no longer available. The moments continue to be defined by the linear set of equations \eqref{Eij}, which remain valid for disconnected, directed graphs. The conjugate gradient iteration needs to be replaced with an iteration for nonsymmetric matrices.

The ability to efficiently calculate moments for every vertex in a large graph is quite powerful, and HITMIX represents just one possible usage of this capability.
We expect that additional techniques can be developed to leverage the information contained in the moment calculations for other purposes, such as outlier detection, community detection, and making inference on global properties of graphs.

\section*{Acknowledgement(s)}

We thank 
Nathan Ellingwood, Julien Rodriguez, Danny Dunlavy and Nate Veldt for assistance and discussions. 
Sandia National Laboratories is a multimission laboratory
managed and operated by National Technology \& Engineering
Solutions of Sandia, LLC, a wholly owned subsidiary of Honeywell
International Inc., for the U.S. Department of Energy’s
National Nuclear Security Administration under contract DENA0003525.

\appendix

\section{Supplementary Tables}

This section includes tables showing results from the figures discussed in Section~\ref{sec:sbm}.

\begin{table}[ht]
\centering
\begin{tabular}{ll|cc}
  \toprule
Method & Number of Blocks & Median $F_1$ & Median ARI \\ 
  \midrule
HITMIX &   2 & 0.95 (0.93, 0.98) & 0.84 (0.74, 0.92) \\ 
  HITMIX &   3 & 0.89 (0.85, 0.93) & 0.63 (0.51, 0.75) \\ 
  HITMIX &   4 & 0.85 (0.80, 0.89) & 0.50 (0.37, 0.62) \\ 
  HITMIX &   5 & 0.82 (0.78, 0.86) & 0.42 (0.31, 0.55) \\ 
  HITMIX &   6 & 0.81 (0.76, 0.85) & 0.39 (0.26, 0.50) \\ 
  HITMIX &   7 & 0.80 (0.76, 0.84) & 0.35 (0.24, 0.47) \\ 
  HITMIX &   8 & 0.79 (0.74, 0.83) & 0.33 (0.23, 0.45) \\ 
  HITMIX &   9 & 0.78 (0.74, 0.82) & 0.32 (0.21, 0.43) \\ 
  HITMIX &  10 & 0.77 (0.73, 0.81) & 0.30 (0.20, 0.40) \\ 
  CRD &   2 & 0.93 (0.90, 0.95) & 0.72 (0.62, 0.81) \\ 
  CRD &   3 & 0.83 (0.71, 0.89) & 0.28 (0.10, 0.47) \\ 
  CRD &   4 & 0.83 (0.65, 0.91) & 0.25 (0.00, 0.46) \\ 
  CRD &   5 & 0.81 (0.24, 0.88) & 0.20 (-0.06, 0.41) \\ 
  CRD &   6 & 0.71 (0.22, 0.85) & 0.03 (-0.06, 0.35) \\ 
  CRD &   7 & 0.41 (0.21, 0.82) & 0.00 (-0.06, 0.30) \\ 
  CRD &   8 & 0.41 (0.21, 0.80) & 0.00 (-0.06, 0.26) \\ 
  CRD &   9 & 0.43 (0.23, 0.77) & 0.00 (-0.06, 0.21) \\ 
  CRD &  10 & 0.44 (0.22, 0.71) & 0.00 (-0.06, 0.09) \\ 
  FS &   2 & 0.18 (0.17, 0.20) & 0.01 (-0.01, 0.03) \\ 
  FS &   3 & 0.18 (0.17, 0.20) & -0.03 (-0.04, -0.02) \\ 
  FS &   4 & 0.18 (0.17, 0.20) & -0.04 (-0.05, -0.04) \\ 
  FS &   5 & 0.18 (0.17, 0.20) & -0.05 (-0.05, -0.04) \\ 
  FS &   6 & 0.18 (0.17, 0.20) & -0.05 (-0.05, -0.04) \\ 
  FS &   7 & 0.18 (0.17, 0.20) & -0.05 (-0.06, -0.05) \\ 
  FS &   8 & 0.18 (0.17, 0.20) & -0.05 (-0.06, -0.05) \\ 
  FS &   9 & 0.18 (0.17, 0.20) & -0.05 (-0.06, -0.05) \\ 
  FS &  10 & 0.18 (0.17, 0.20) & -0.06 (-0.06, -0.05) \\ 
  HK &   2 & 0.62 (0.56, 0.67) & 0.19 (0.11, 0.28) \\ 
  HK &   3 & 0.55 (0.50, 0.59) & -0.01 (-0.04, 0.03) \\ 
  HK &   4 & 0.52 (0.47, 0.56) & -0.05 (-0.07, -0.03) \\ 
  HK &   5 & 0.50 (0.45, 0.54) & -0.07 (-0.08, -0.05) \\ 
  HK &   6 & 0.49 (0.44, 0.54) & -0.07 (-0.09, -0.06) \\ 
  HK &   7 & 0.48 (0.44, 0.52) & -0.08 (-0.09, -0.07) \\ 
  HK &   8 & 0.47 (0.43, 0.52) & -0.08 (-0.09, -0.07) \\ 
  HK &   9 & 0.47 (0.43, 0.51) & -0.08 (-0.10, -0.07) \\ 
  HK &  10 & 0.47 (0.43, 0.51) & -0.08 (-0.10, -0.07) \\ 
  SL &   2 & 0.17 (0.14, 0.19) & 0.01 (-0.01, 0.03) \\ 
  SL &   3 & 0.18 (0.15, 0.19) & -0.03 (-0.04, -0.02) \\ 
  SL &   4 & 0.18 (0.15, 0.19) & -0.04 (-0.05, -0.03) \\ 
  SL &   5 & 0.18 (0.16, 0.19) & -0.05 (-0.05, -0.04) \\ 
  SL &   6 & 0.18 (0.15, 0.19) & -0.05 (-0.05, -0.04) \\ 
  SL &   7 & 0.18 (0.16, 0.19) & -0.05 (-0.06, -0.04) \\ 
  SL &   8 & 0.18 (0.15, 0.19) & -0.05 (-0.06, -0.05) \\ 
  SL &   9 & 0.18 (0.16, 0.20) & -0.05 (-0.06, -0.05) \\ 
  SL &  10 & 0.18 (0.15, 0.19) & -0.05 (-0.06, -0.05) \\ 
   \bottomrule
\end{tabular}
\caption{$F_1$ and ARI scores for various values of number of blocks in SBM simulation 1. Parenthetical values are 5 and 95\% percentiles. Percentiles are taken across Monte Carlo Samples. 500 Monte Carlo samples were taken per setting of number of blocks.} 
\label{tab:sbmBlock}
\end{table}

\begin{table}[ht]
\centering
\begin{tabular}{ll|cc}
  \toprule
Method & $p_{in}$ & Median $F_1$ & Median ARI \\ 
  \midrule
HITMIX & 0.20 & 0.98 (0.96, 1.00) & 0.94 (0.86, 1.00) \\ 
  HITMIX & 0.18 & 0.96 (0.93, 0.99) & 0.88 (0.74, 0.96) \\ 
  HITMIX & 0.16 & 0.93 (0.88, 0.96) & 0.74 (0.59, 0.88) \\ 
  HITMIX & 0.14 & 0.86 (0.78, 0.92) & 0.57 (0.38, 0.74) \\ 
  HITMIX & 0.12 & 0.77 (0.67, 0.84) & 0.37 (0.21, 0.53) \\ 
  HITMIX & 0.10 & 0.66 (0.56, 0.75) & 0.19 (0.08, 0.33) \\ 
  HITMIX & 0.08 & 0.55 (0.44, 0.65) & 0.06 (0.01, 0.15) \\ 
  HITMIX & 0.06 & 0.45 (0.35, 0.55) & 0.00 (-0.01, 0.04) \\ 
  CRD & 0.20 & 0.98 (0.92, 1.00) & 0.90 (0.69, 1.00) \\ 
  CRD & 0.18 & 0.93 (0.88, 0.99) & 0.74 (0.56, 0.96) \\ 
  CRD & 0.16 & 0.89 (0.84, 0.94) & 0.59 (0.43, 0.78) \\ 
  CRD & 0.14 & 0.84 (0.78, 0.90) & 0.46 (0.30, 0.62) \\ 
  CRD & 0.12 & 0.78 (0.71, 0.84) & 0.31 (0.18, 0.45) \\ 
  CRD & 0.10 & 0.72 (0.64, 0.78) & 0.18 (0.08, 0.29) \\ 
  CRD & 0.08 & 0.65 (0.56, 0.71) & 0.07 (0.02, 0.16) \\ 
  CRD & 0.06 & 0.57 (0.48, 0.64) & 0.01 (0.00, 0.06) \\ 
  FS & 0.20 & 0.18 (0.17, 0.20) & 0.01 (-0.01, 0.03) \\ 
  FS & 0.18 & 0.18 (0.16, 0.20) & 0.01 (-0.01, 0.04) \\ 
  FS & 0.16 & 0.18 (0.17, 0.20) & 0.01 (-0.01, 0.04) \\ 
  FS & 0.14 & 0.18 (0.17, 0.21) & 0.01 (-0.01, 0.04) \\ 
  FS & 0.12 & 0.18 (0.17, 0.21) & 0.01 (-0.01, 0.04) \\ 
  FS & 0.10 & 0.18 (0.17, 0.21) & 0.01 (-0.01, 0.04) \\ 
  FS & 0.08 & 0.18 (0.16, 0.20) & 0.01 (-0.01, 0.03) \\ 
  FS & 0.06 & 0.18 (0.16, 0.20) & 0.01 (-0.01, 0.03) \\ 
  HK & 0.20 & 0.68 (0.60, 0.76) & 0.25 (0.12, 0.41) \\ 
  HK & 0.18 & 0.68 (0.61, 0.77) & 0.25 (0.12, 0.42) \\ 
  HK & 0.16 & 0.69 (0.61, 0.75) & 0.24 (0.12, 0.37) \\ 
  HK & 0.14 & 0.68 (0.62, 0.75) & 0.21 (0.11, 0.35) \\ 
  HK & 0.12 & 0.66 (0.60, 0.72) & 0.15 (0.06, 0.26) \\ 
  HK & 0.10 & 0.63 (0.57, 0.70) & 0.09 (0.03, 0.18) \\ 
  HK & 0.08 & 0.61 (0.54, 0.67) & 0.04 (0.00, 0.11) \\ 
  HK & 0.06 & 0.56 (0.51, 0.63) & 0.01 (0.00, 0.05) \\ 
  SL & 0.20 & 0.17 (0.12, 0.19) & 0.01 (-0.01, 0.03) \\ 
  SL & 0.18 & 0.16 (0.11, 0.20) & 0.01 (-0.01, 0.03) \\ 
  SL & 0.16 & 0.15 (0.09, 0.19) & 0.01 (-0.01, 0.03) \\ 
  SL & 0.14 & 0.14 (0.09, 0.19) & 0.01 (-0.01, 0.03) \\ 
  SL & 0.12 & 0.13 (0.07, 0.18) & 0.00 (-0.01, 0.03) \\ 
  SL & 0.10 & 0.12 (0.06, 0.17) & 0.00 (-0.01, 0.02) \\ 
  SL & 0.08 & 0.10 (0.05, 0.16) & 0.00 (-0.01, 0.02) \\ 
  SL & 0.06 & 0.08 (0.04, 0.14) & 0.00 (-0.01, 0.01) \\ 
   \bottomrule
\end{tabular}
\caption{$F_1$ and ARI scores for a subset of values of $p_{in}$ in SBM simulation 2. Parenthetical values are 5 and 95\% percentiles. Percentiles are taken across Monte Carlo Samples. 500 Monte Carlo samples were taken per setting of $p_{in}$.} 
\label{tab:sbmPin}
\end{table}

\begin{table}[ht]
\centering
\begin{tabular}{ll|cc}
  \toprule
Method & Size of Hitting Set & Median $F_1$ & Median ARI \\ 
  \midrule
HITMIX & 50 & 1.00 (0.98, 1.00) & 1.00 (0.95, 1.00) \\ 
  HITMIX & 25 & 0.98 (0.95, 0.99) & 0.93 (0.82, 0.98) \\ 
  HITMIX & 10 & 0.89 (0.83, 0.95) & 0.66 (0.47, 0.82) \\ 
  HITMIX & 5 & 0.78 (0.66, 0.85) & 0.37 (0.20, 0.55) \\ 
  HITMIX & 1 & 0.38 (0.19, 0.60) & 0.02 (-0.01, 0.09) \\ 
  CRD & 50 & 0.69 (0.53, 1.00) & 0.30 (0.00, 0.98) \\ 
  CRD & 25 & 0.62 (0.42, 0.97) & 0.20 (0.00, 0.90) \\ 
  CRD & 10 & 0.87 (0.81, 0.92) & 0.53 (0.36, 0.72) \\ 
  CRD & 5 & 0.80 (0.72, 0.89) & 0.36 (0.20, 0.63) \\ 
  CRD & 1 & 0.60 (0.42, 0.80) & 0.13 (0.00, 0.39) \\ 
  FS & 50 & 0.67 (0.62, 0.73) & 0.25 (0.15, 0.38) \\ 
  FS & 25 & 0.40 (0.37, 0.44) & 0.06 (0.01, 0.12) \\ 
  FS & 10 & 0.18 (0.16, 0.20) & 0.01 (-0.01, 0.04) \\ 
  FS & 5 & 0.10 (0.09, 0.11) & 0.00 (-0.01, 0.02) \\ 
  FS & 1 & 0.02 (0.02, 0.02) & 0.00 (0.00, 0.00) \\ 
  HK & 50 & 0.77 (0.70, 0.84) & 0.38 (0.23, 0.56) \\ 
  HK & 25 & 0.74 (0.67, 0.82) & 0.33 (0.20, 0.52) \\ 
  HK & 10 & 0.69 (0.62, 0.75) & 0.23 (0.11, 0.35) \\ 
  HK & 5 & 0.61 (0.55, 0.68) & 0.11 (0.04, 0.22) \\ 
  HK & 1 & 0.62 (0.51, 0.71) & 0.13 (0.02, 0.29) \\ 
  SL & 50 & 0.67 (0.62, 0.72) & 0.25 (0.14, 0.37) \\ 
  SL & 25 & 0.39 (0.34, 0.43) & 0.06 (0.01, 0.12) \\ 
  SL & 10 & 0.15 (0.09, 0.19) & 0.01 (-0.01, 0.03) \\ 
  SL & 5 & 0.07 (0.04, 0.10) & 0.00 (-0.01, 0.01) \\ 
  SL & 1 & 0.02 (0.02, 0.02) & 0.00 (0.00, 0.00) \\ 
   \bottomrule
\end{tabular}
\caption{$F_1$ and ARI scores for various hitting set sizes in SBM simulation 3. Parenthetical values are 5 and 95\% percentiles. Percentiles are taken across Monte Carlo Samples. 500 Monte Carlo samples were taken per setting of hitting set size.} 
\label{tab:sbmHit}
\end{table}

\clearpage

\section*{Funding}
This paper describes objective technical results and analysis. Any subjective views or opinions that might be expressed in the paper do not necessarily represent the views of the U.S. Department of Energy or the United States Government.
Sandia National Laboratories, a multimission laboratory managed and operated by National Technology and Engineering Solutions of
Sandia, LLC., a wholly owned subsidiary of Honeywell International, Inc., for the U.S. Department of Energy National Nuclear Administration under contract DE-NA0003525.
SAND2020-12836O.

\bibliographystyle{tfs}
\bibliography{network}

\appendix

\end{document}